# Electrochemical solid-state amorphization in the immiscible Cu-Li system: Size matters

*Muhua Sun[1,2], Jiake Wei[1], Zhi Xu[1,2], Qianming Huang[1,2], Yu Zhao[1], Wenlong Wang[1,2]\*, Xuedong Bai[1,2]\**

[1]Beijing National Laboratory for Condensed Matter Physics and Institute of Physics, Chinese Academy of Sciences, Beijing 100190, China

[2]School of Physical Sciences, University of Chinese Academy of Sciences, Beijing 100190, China

Abstract:

As a typical immiscible binary system, copper (Cu) and lithium (Li) show no alloying and chemical intermixing under normal circumstances. Here we show that, when decreasing Cu nanoparticle sizes into ultrasmall range, the nanoscale size effect can play a subtle yet critical role in mediating the chemical activity of Cu and therefore its miscibility with Li, such that the electrochemical alloying and solid-state amorphization will occur in such an immiscible system. This unusual observation was accomplished by performing in-situ studies of the electrochemical lithiation processes of individual CuO nanowires inside a transmission electron microscopy (TEM). Upon lithiation, CuO nanowires are first electrochemically reduced to form discrete ultrasmall Cu nanocrystals that, unexpectedly, can in turn undergo further electrochemical lithiation to form amorphous $CuLi_x$ nanoalloys. Real-time TEM imaging unveils that there is a critical grain size (ca. 6 nm), below which the nanocrystalline

Cu particles can be continuously lithiated and amorphized. The possibility that the observed solid-state amorphization of Cu-Li might be induced by electron beam irradiation effect can be explicitly ruled out; on the contrary, it was found that electron beam irradiation will lead to the dealloying of as-formed amorphous $CuLi_x$ nanoalloys.

Since the first discovery of amorphous metallic alloys in the early of 1960s, the past half a century has witnessed tremendous growth in both synthetic sophistication and depth characterization of this important class of materials.[1] Conventionally, amorphous metallic alloys can be synthesized by the rapid solidification of molten alloys, which is basically a physical method. As a fundamentally different scenario, the chemically driven solid-state amorphization that relies on the diffusive solid-state reactions at moderate temperature between pure metals has also been demonstrated early in 1983 as a viable approach for amorphous alloys formation.[2] More recently, it has been found that such a diffusive reaction based solid-state amorphization process can also be driven by electrochemical alloying at room temperature.[3]

The most studied systems to date in the context of electrochemical solid-state amorphization are Li-Si and Li-Sn amorphous alloys.[4] Primary interest in these two systems stems from the potential utilization of Si and Sn as high-capacity anode materials in Li-ion batteries. Both Li-Si and Li-Sn are typical miscible binary systems, with a negative heat of mixing ($\Delta H^{mix}$) as large as -30 kJ/mol and -18 kJ/mol,[5] respectively. This large negative $\Delta H^{mix}$ provides the thermodynamic basis to enable chemical intermixing and amorphization through the diffusive solid-state reactions between Si (or Sn) and electrochemically generated Li.[6] In sharp contrast, however, Cu-Li is a well-known immiscible system with their $\Delta H^{mix}$ being near zero (0.56 kJ/mol by Calphad approach and -5 kJ/mol calculated by Miedema's model).[5, 7] As such, Cu and Li have no or little tendency to spontaneously alloy on atomic scale in thermodynamic equilibrium.[8] A notable example that takes advantages of the immiscibility between Cu and Li is the widespread utilization of Cu foils as the anodic current collector in Li-ion batteries. Owning to the equilibrium-immiscible nature of Cu with Li, the electrochemically generated Li atoms show no intermixing and alloying with the Cu current collectors (Figure 1a), thereby assuring the long-term operation stability of Li-ion batteries.[9] Here we show for the first time that the nanoscale size effect can play a subtle yet critical role in mediating the chemical activity of Cu and therefore its miscibility with Li. When decreasing nanoparticle sizes into ultrasmall range, Cu as a parent phase can react readily with the electrochemically generated Li, thereby leading to the occurrence of diffusion alloying and solid-state amorphization in such an equilibrium-immiscible binary metallic system (Figure 1b).

Our work was accomplished by taking advantage of an in-situ transmission electron microscopy (TEM) platform that allows the real-time studies of nanoscale electrochemical cells under dynamic operating conditions. Experimentally, a solid-state open cell configuration (Figure S1) was adopted for in-situ TEM studies, with the selected individual CuO nanowire (fixed on a gold tip) as the observable electrode, bulk Li metal (fixed on the tungsten tip) as the counter electrode, and the naturally grown thin $Li_2O$ layer on Li metal as the solid-state electrolyte.[10] Upon lithiation, CuO nanowires were electrochemically reduced to discrete ultrasmall Cu nanoparticles that are embedded in $Li_2O$ matrix (Figure S4), as consistent with that reported earlier.[11] Here we denote this lithiation process ($\mathbf{CuO + Li^+ + e^- \rightarrow Cu + Li_2O}$) as the "1st stage" lithiation. An intriguing new finding in our present study is the observation that, following the "1st stage" lithiation process, there will occur a previously unknown "2nd stage" electrochemical lithiation process, i.e., the as-formed ultrasmall Cu nanoparticles embedded in $Li_2O$ matrix can be further electrochemically lithiated to form amorphous $CuLi_x$ nanoalloys ($\mathbf{Cu + Li^+ + e^- \rightarrow CuLi_x}$).

In Figure 2, we show how the two stages of lithiation processes occur in sequence. Upon the "1st stage" lithiation of CuO nanowire, a clear reaction front that separates the lithiated phases from the pristine CuO phase can be clearly seen in the nanowire (Figure 2a). With the occurrence of "2nd stage" lithiation, a new reaction front will emerge, along with the first one; as a result, the nanowire is separated into three distinct segments (Figure 2b). A continuous movie that recording the entire lithiation processes is shown in Supporting Information (movie S1). Interestingly, it can be noticed that the newly formed "2nd stage" lithiated region exhibits a much lighter image contrast, which is reminiscent of the possible formation of some amorphous phases. To confirm this, we comparatively probed these two lithiated regions with selected area electron diffraction (SAED). Whereas SAED pattern of the "1st stage" lithiated region shows the clear presence of diffraction rings both from $Li_2O$ and Cu, SAED pattern of the "2nd stage" lithiated region displays only the diffraction rings of $Li_2O$, as expected. Taken together, the observed distinct contrast fading of TEM image and the disappearance of Cu diffraction rings thereby confirm that the nanocrystalline Cu particles were amorphized during the "2nd stage" lithiation process.

    A key question then arises: is this amorphization event caused by simple crystalline-to-amorphous transformation of Cu nanoparticles themselves, or otherwise, by lithiation-induced solid-state amorphization of nanocrystalline Cu particles, just like what happens in Li-Si and Li-Sn systems. In our present work, we have got multiple complementary experimental evidences to validate that this amorphization event is due to the latter case. The first evidence comes from Z-contrast high-angle annular dark field (HAADF) images where the formation of Cu-Li alloying phase can be identified. Electron energy loss spectroscopy (EELS) characterization further reveals that there is a net charge transfer from Li to Cu, a straightforward evidence of the alloying between Cu and Li. Moreover, real-time kinetic measurements of the "2nd stage" lithiation process also unveil a very slow reaction rate that is typical of the solid-state diffusive alloying reactions. The last evidence comes from a remarkable finding that the deliberate irradiation of the as-formed amorphous $CuLi_x$

nanoalloys by electron beam can induce the delithiation of them, i.e., the amorphous alloying phenomena is reversible in nature.

Figure 3a displays a representative HAADF image that depicts a comparison between the "1st stage" and "2nd stage" lithiated regions in a same nanowire. Because the atomic number of Cu is far larger than that of Li and the image intensity of HAADF is known to be approximately proportional to the square of atomic number ($Z^2$), what is highlighted in Figure 3a is mainly the Cu element whereas Li element (in its any forms) is hardly visible. As such, the observed reduced HAADF image intensity in the "2nd stage" lithiated region indicates the formation of a "diluted Cu" phase, i.e., the ultrasmall Cu nanoparticles are lithiated to form the $CuLi_x$ alloying phase. As a complementary evidence of HAADF, we show in Figure 3b and 3c the corresponding bright-field high resolution TEM (HRTEM) images of the two different lithiated regions. In the "1st stage" lithiated region (Figure 3b), ultrasmall Cu nanoparticles can be clearly discerned from the crystalline $Li_2O$ matrix. Whereas in the "2nd stage" lithiated region, HRTEM only shows the lattice fringes of $Li_2O$ matrix, demonstrating that Cu-Li alloying phase is of an amorphous nature. This result is in well consistence with the SAED results as discussed above.

Figure 3d shows a comparison of Cu-$L_{2,3}$ EELS spectra of CuO, the ultramall Cu nanoparticles dispersed in $Li_2O$ matrix and the amorphous $CuLi_x$ alloys, respectively. For transition metals, $L_{2,3}$ edge arises through dipole transition from core-level 2p electrons to the narrow unoccupied d states. Normally, two sharp peaks known as "white lines" can be observed at the onsets of the $L_2$ and $L_3$ absorption edges, the intensities of which show strong correlations with d-state occupancy.[12] In the case of elemental Cu, where there are no unfilled d states, no sharp white lines can be observed, leaving only steps in $L_{2,3}$ absorption edge.[13] Upon oxidation to form CuO, the electron transfer from the Cu 3d orbital to oxygen will lead to the reappearance of white lines and the lowered EELS threshold energy of the $L_3$ edge compared to Upon oxidation to form CuO, the electron transfer from the Cu 3d orbital to oxygen will lead to the appearance of white lines together with a slightly lowered EELS threshold energy of the $L_3$ edge, as seen in Figure 3d. As far as Cu-based alloys are concerned, the situation is much more complicated. When alloying with early transition metals with low d-state occupancy, such as in Cu-Zr and Cu-Ti, EELS spectra are featured by a large enhancement of the white line intensities, indicating that there is a net d-electron transfer from Cu to Zr or Ti.[14] On the contrary, when alloying of Cu with typical electron-donating metals, such as Cu-Li system in our case, one can expect that charge transfer will occur from Li to Cu, which will increase the overall occupancy of Cu 3d-4s band and thereby diminish the intensity of $L_{2,3}$ absorption edge. As shown in Figure 3d, the spectrum feature of amorphous Cu-Li nanoalloys is well consistent with these expectations, although it is difficult to quantify the amount of transferred charge.

As compared to the "1st stage" lithiation, the "2nd stage" lithiation proceeds much slower. This is understandable, considering that the solid-state Cu-Li alloying reaction involves a sluggish diffusion process, that is, the slow diffusion of the freshly formed Li atoms into the lattice of pre-formed Cu nanoparticles.[15] From the time-lapse TEM images in Figure 4a, we can clearly see the slow propagation of the reaction front during the "2nd stage" lithiation process. In-situ kinetic measurement was made, and a plot of L (the moving distance of the "2nd front" compared to the position of it at the selected time of 0 s) versus time t, is shown in Figure 4b where the deduced reaction rate ≈ 0.3 nm/s. It is important to note that such a slow reaction rate is common for all of the tested nanowire samples (Figure S6), although the deduced values vary a little from sample to sample. As a whole, the reaction rate of "2nd stage" lithiation is around one order of magnitude slower than that of the "1st stage" lithiation.[11a] Moreover, the nearly perfect linear relationship between L and t (Figure 4b and Figure S6) implies that the electrochemical alloying of Cu-Li is a reaction limited process. In other words, the slow diffusive Cu-Li alloying reaction that occurs at a rate limited by Li diffusion in lattice of Cu nanoparticles, is a decisive step over the "2nd stage" lithiation process.

A key determinant of solid-state amorphization, either chemically or electrochemically, is to elevate the free energy of the pure reactants to an energy level being higher than that of the metastable amorphous alloying phases. For systems with a large negative heat of mixing, such as Li-Si and Li-Sn, this thermodynamic requirement is relatively easier to fulfil, and therefore the solid-state amorphization reaction can more readily occur. By contrast, for an immiscible system with less negative or even positive heat of mixing, such as Cu-Li, the occurrence of solid-state alloying and amorphization is less likely under normal circumstances. Nevertheless, by purposely utilizing the nanoscale size effects to impart excess surface/interface free energy, the free energy of reactants can be elevated to high "ladder" such that the amorphous alloying becomes possible.[16] In our present work, the Cu nanoparticles resulting from the in-situ reduction of CuO nanowires during "1st stage" lithiation, have got an ultrasmall size region focusing around 2-5 nm (Figure 4c). This extremely ultrasmall size marks the point at which the majority of atoms in a specific Cu nanoparticle are at its surface,[17] which can dramatically increase the surface free energy of the system compared to bulk Cu or larger crystal size Cu.[8a, 18] Therefore, it is the nanoscale size effect that plays a subtle yet critical role in mediating the chemical activity of Cu and therefore its miscibility with Li, which allows for the occurrence of electrochemical alloying and amorphization in such an immiscible system. Interestingly, with a close examination of the time-lapse images in Figure 4a (also the HAADF image in Figure 3a) a fact that should be not ignored is that, although the large majority of Cu nanoparticles were transformed to amorphous CuLi$_x$ nanoalloys, there still are some larger-sized Cu nanoparticles being left behind the reaction front. This implies that there exists a critical size above which the nanocrystalline Cu particles **cannot** be lithiated to form amorphous phases, a key manifestation of the nanoscale size effect. From the comparative statistic results shown in Figure 4c, the critical grain size can be experimentally identified to be ca. 6 nm. Moreover, we also carried out control experiments with the purposely synthesized larger Cu nanoparticles (Figure S7) and single crystalline Cu nanowires (Figure S7) to verify the nanoscale size effect and existence of critical size.

For in-situ TEM studies, the effect of the imaging electron beam irradiation requires careful evaluation. In our present work, we have carefully examined whether or not there is possibility that the occurrence of Cu-Li amorphization was induced by the electron beam irradiation. When the electron beam was deliberately blanked or spread to be extremely weak (Figure S9), it was found that the solid-state Cu-Li alloying and amorphization remained to occur. Therefore, it is safe to say that this process is exactly electrochemically driven, rather than due to electron beam irradiation effect. On the contrary, it was surprisingly found that the electron beam irradiation could actually induce an opposite result. That is, under intense electron beam irradiation, the as-formed amorphous $CuLi_x$ alloys would be delithiated, accompanied with the transformation of amorphous $CuLi_x$ alloys to metallic Cu nanoparticles (Figure S10). Here it is worth noting that the similar electron-beam-induced-delithiation phenomenon had also been previously observed by Wang et. al when studying the lithiation of Si nanowire.[19] Although the underlying mechanism of the electron beam induced dealloying phenomenon remains elusive, this remarkable observation anyhow demonstrates that the possibility that the observed solid- state amorphization of Cu-Li might be induced by electron beam irradiation effect can be explicitly ruled out.

To sum up, by taking advantage of in-situ TEM dynamic observations, the electrochemical alloying and amorphization in immiscible Cu-Li system is clearly revealed. Given the widespread utilization of Cu as the anodic current collector in commercial Li-ion batteries, our work may reveal a previously unknown, yet important, consideration that accounts for the capacity fade and current collector degradation problems during battery operation. In broader terms, our experimental finding of the distinct nanoscale size effect that mediates the binary Cu-Li system from immiscible to miscible, also reveals an important new insight for fundamental understanding of the diffusive solid-state reactions in general.

**Acknowledgements**
This work was supported by the Program from Chinese Academy of Sciences (Grant Nos. ZDYZ2015-1 and XDB07030100), Natural Science Foundation (Grant Nos. 11474337, 51421002, 51472267, 51172273, and 221322304) of China and Austrian-Chinese Cooperative R&D Projects, FFG and CAS, No. 112111KYSB20150002.

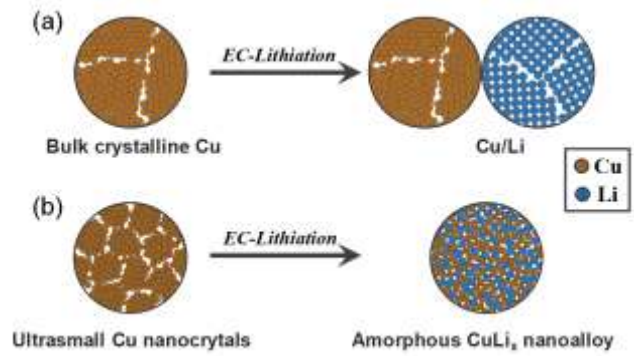

**Figure 1.** A schematic of the immiscibility of Cu towards electrochemically generated Li. (a) Bulk Cu shows no alloying with Li. (b) When decreasing Cu nanoparticle sizes into ultrasmall region, the electrochemical (EC) lithiation of Cu will occur and lead to the formation of amorphous $CuLi_x$ nanoalloys.

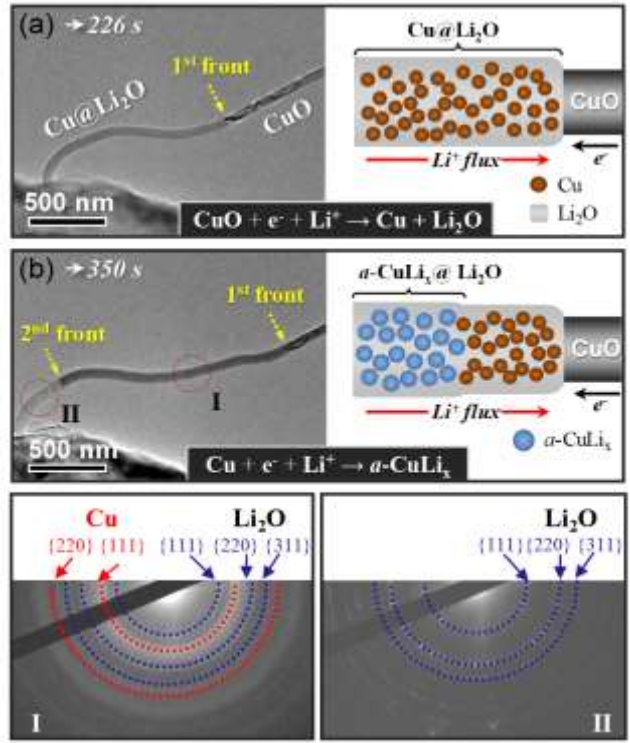

**Figure 2.** In-situ TEM observation of the two-stage electrochemical lithiation processes. (a) TEM image and the schematic illustration of CuO nanowire after "1st stage" lithiation. (b) The same CuO nanowire imaged upon the occurrence of "2nd stage" lithiation, showing the simultaneous presence of two reaction fronts. The two panels at the bottom display SAED patterns of the two different lithiated areas, as marked in (b).

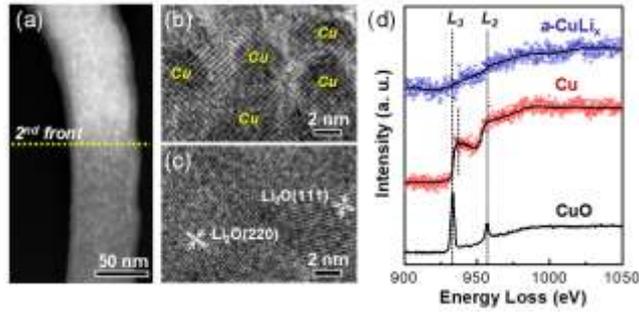

**Figure 3.** (a) HAADF image of the two lithiated regions across the second reaction front. (b)-(c) HRTEM images of the "1st stage" and "2nd stage" lithiated regions, respectively. (d) A comparison of EELS spectra of Cu $L_{2,3}$ edge collected from pristine CuO nanowire, Cu nanoparticles embedded in $Li_2O$ matrix, and amorphous $CuLi_x$ nanoalloys embedded in $Li_2O$ matrix.

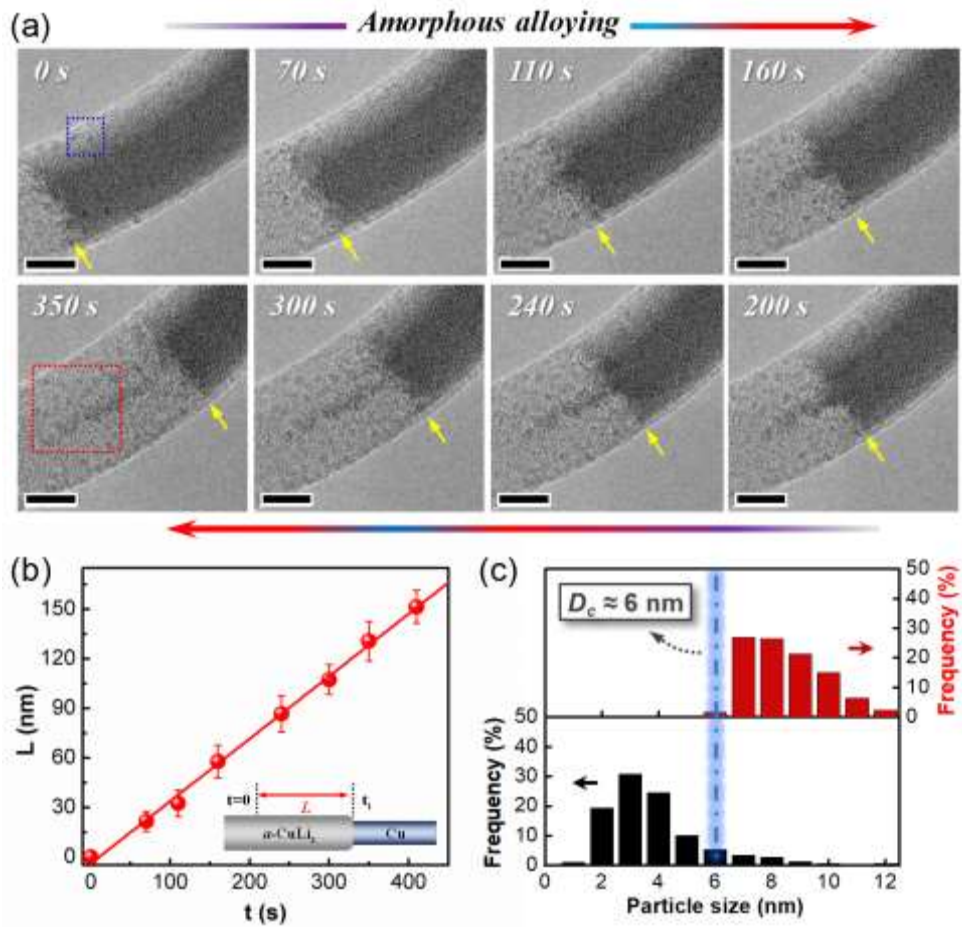

**Figure 4.** Lithiation kinetics and nanoscale size effect in Cu-Li electrochemical alloying and amorphization. (a) Time-lapse TEM images showing the propagation of the reaction front during "2nd stage" lithiation (See also Movie S2 in Supporting Information). The yellow arrow in each frame indicates the "2nd front". scale bar, 50 nm. (b) A plot of distance L (the "2nd stage" reaction front propagation length) versus the corresponding time t. (c) The statistical particle size distribution of the unalloyed Cu nanoparticles (top) and the pristine as-formed Cu nanoparticles (bottom), correspondingly measured from the area framed in (a) by the red rectangle and the HRTEM image in figure S4 of the blue rectangle framed area